\documentclass[usenatbib,usegraphicx]{mn2e}

\begin{document}

\title[Very high contrast IFU spectroscopy]
{Very high contrast IFU spectroscopy of AB Doradus C:
9 mag contrast at 0\farcs 2 without a coronagraph using spectral deconvolution.
\thanks{Based on observations collected at the European
  Southern Observatory, Chile, under ESO programme 276.C-5013.}}

\author[Niranjan Thatte et al.]{Niranjan Thatte$^1$, 
  Roberto Abuter$^2$, Matthias Tecza$^1$, Eric L.~Nielsen$^3$,
\newauthor Fraser J.~Clarke$^1$, \& Laird M.~Close$^3$\\
$^1$ Department of Astrophysics, Denys Wilkinson Building, Keble Road, Oxford, OX1 3RH.\\
$^2$ Max-Planck-Institut f\"ur extraterrestrische Physik \& European Southern Observatory, D-85748, Garching, Germany.\\
$^3$ Steward Observatory, University of Arizona, Tucson, AZ 85721,
  U.S.A.\\
email: nthatte@astro.ox.ac.uk (NT), rabuter@eso.org (RA),
  mtecza@astro.ox.ac.uk (MT), enielsen@as.arizona.edu (EN),\\ 
fclarke@astro.ox.ac.uk (FJC), lclose@as.arizona.edu (LMC)}

\date{accepted 2007 March 8, received 2007 March 8, in original form 2006 June 6}

\pagerange{\pageref{firstpage}--\pageref{lastpage}} \pubyear{2007}

\maketitle

\label{firstpage}

\begin{abstract}
We present an extension of the spectral deconvolution method
\citep{sparks02} to achieve very high contrast at small inner
working radii. We apply the method to the specific case of ground
based adaptive optics fed integral field spectroscopy (without a
coronagraph).  Utilising the wavelength dependence of the Airy and
speckle patterns, we make an accurate estimate of the PSF that can be
scaled and subtracted from the data cube. The residual noise in the
resulting spectra is very close to the photon noise from the starlight
halo. We utilise the technique to extract a very high SNR H \& K band
spectrum of AB Dor C, the low mass companion to AB Dor A. By
effectively eliminating all contamination from AB Dor A, the extracted
spectrum retains both continuum and spectral features. The achieved
1$\sigma$ contrast is 9 mag at 0\farcs 2, 11 mag at 0\farcs 5, in 20
mins exposure time, at an effective spectral bandwidth of 5.5 nm,
proving that the method is applicable even in low Strehl regimes.
 
The spectral deconvolution method clearly demonstrates the efficacy of
image slicer based IFUs in achieving very high contrast imaging
spectroscopy at small angular separations, validating their use as
high contrast spectrographs/imagers for extreme adaptive optics
systems.
\end{abstract}

\section{Introduction}

Direct imaging and spectroscopy of extra-solar planets is a research
area that has attracted significant attention in recent years. Several
specialised instruments are now being designed and built to achieve
this goal within the next few years (e.g. VLT Planet Finder (SPHERE)
-- \citet{sphere06}, Gemini GPI -- \citet{gpi06}, Keck XAOPI -- 
\citet{macintosh04}). The major difficulty in detecting exo-planets is
the extreme contrast between the star (typically a main-sequence M--G
dwarf) and the planet (typically less than 1\arcsec\, away). Even in
the most favourable conditions, theoretical computations
\citep[e.g.][]{burrows04} have shown that the brightness ratio between
the parent star and the exo-planet is expected to be at least 15
magnitudes at 1\arcsec.  Such contrasts are significantly beyond
normal imaging techniques even with the best adaptive optics systems
available today. If we are to attain this ambitious goal, novel ways
to maximise the achieved contrast must be devised.

Even in high-Strehl AO systems, long exposures do not provide a smooth
halo around the central star, as one would expect from a time average
of the speckles caused by atmospheric turbulence. In fact, due to the
presence of super-speckles, imaging contrasts are much worse than the
theoretical {\em photon noise} limit \citep{racine99}. Super-speckles
(also called quasi-static speckles \citet{adi06})
are long lived speckles caused by imperfections in the light path, and
have coherence times ranging from tens of seconds to several
minutes. Any changes in the light path, for example due to instrument
flexure or changing telescope position, cause the super-speckles to
vary, making it impossible to remove them with static calibrations
obtained at a different time (e.g. PSF star observations will show a
completely different super-speckle pattern).

The contrast limitations posed by super-speckles can only be overcome
through {\em simultaneous} observations. Several techniques have been
proposed to achieve this \citep[e.g.][]{guyon04,marois04,adi06,ren06}
including the ``spectral difference imager'' (SDI) technique developed
by one of the authors \citep{lenzen04} and others \citep{trident05}.
The technique utilises the fact that the planetary spectrum has sharp,
deep absorption features (e.g. CH$_4$ in the H band) to dramatically
improve the contrast with respect to scattered starlight, which has a
smooth spectrum.  By taking simultaneous images in and out of the
methane absorption band, and scaling and subtracting the two images,
one effectively removes the starlight, while leaving the planetary
light intact. The subtraction removes residual starlight, whether in
the form of rapidly time-varying speckles, slowly varying
super-speckles, or any other form of scattered light within the
telescope \& instrument.  In fact, the SDI technique uses an ingenious
double-difference, so as to further reduce any residuals arising from
systematic errors due to differing optical paths. \citet{biller06} have
demonstrated the success of SDI in finding cool companions with the
NACO instrument on the ESO-VLT.

SDI is however limited in application as it relies on an intrinsic
feature of the companion spectrum; the CH$_4$ absorption feature only
found in objects with T$_{\rm eff} < $ 1200\,K. In addition, there is a
need for follow-up spectroscopy of the candidate object detected --
both to confirm its nature (often via common proper motion with the
parent), and to characterise it in detail. In this paper, we describe
an extension 
of the \citet{sparks02} {\em spectral deconvolution} (SD) technique 
to achieve high-contrast imaging (and spectroscopy) with
an integral field spectrograph (IFS). This technique does not rely on
any feature of the target object's spectrum and is therefore
applicable to any high contrast imaging application. In addition, the
use of an IFS provides a complete dataset in one go, and with maximum
signal-to-noise (no slit losses). Furthermore, complete 2D spatial
information allows the continuum to be correctly measured, in contrast
to extremely narrow slits needed for A.O. assisted spectroscopy. No
a-priori information about the spatial location of the companion is
needed either, an inherent advantage of using an IFS.

The next two section describes the \citet{sparks02} method, as well as
proposed extensions, including using this technique at small inner
working radii (within the bifurcation radius, see section
\ref{bifurcation}), and in section \ref{demo} we provide a
demonstration of the achievable performance in this regime, {\em even
without a coronagraph}. Section \ref{results} presents the results and
section \ref{conclusion} elaborates on the future prospects and the
discovery potential for high contrast imaging spectroscopy with image
slicer based integral field spectrographs, utilising this technique.

\section{Concept}\label{concept}

Several authors \citep[e.g.][]{sparks02,fusco05,berton06b} have
suggested that the wealth of spectral information available in an IFS
data cube can be utilised to remove scattered starlight and identify
the presence of a close-in companion, and extract its spectrum with
enhanced signal-to-noise (SNR), thus maximising contrast. Our scheme is an
extension of the spectral deconvolution technique proposed by
\citet{sparks02}, adapted to ground based observing with both extreme
adaptive optics and general purpose adaptive optics systems, and further
optimised to maximise SNR.  Application of the concept to ground based
IFS fed by general purpose A.O. systems is very challenging, as the
Strehl ratio is typically only $\sim$0.5, and varies strongly with
wavelength. 

The central tenet of the SD technique is that the position of almost
all features arising from the parent star (Airy pattern, speckles,
etc.)  scale smoothly and slowly with wavelength (e.g. the first Airy
null is always 1.22\,$\lambda$/D from the star), whereas a physical
object's location is independent of wavelength. Thus, stepping through
the wavelength axis of an IFS data cube at a fixed spatial
location (e.g. the location of a companion), one would see a
modulation arising from maxima and minima of the Airy pattern passing
through the companion's location \citep[see figure 24
in][]{sparks02}. These modulations would drown out any expected signal
from the companion. If however, we use the wavelength dependence to
advantage, we can effectively subtract all wavelength dependent
artifacts in the stellar point spread function (PSF), thus unmasking
the presence of real physical objects.

The technique is best illustrated via the removal of the Airy pattern
from the IFS data cube, although it applies equally well to any {\em
achromatic} aberration in the wavefront. Each wavelength slice of the
IFS data cube is scaled radially, so as to exactly compensate the
wavelength dependence of the Airy pattern.  Nulls and maxima in the
Airy pattern, which used to be located at different radial distances
in the different wavelength slices now line up perfectly at all
wavelengths in the scaled data cube \citep[e.g. figure 27
in][]{sparks02}. The companion's light now traces a diagonal line
through the data cube.

\citet{sparks02} fit a low order polynomial to every spatial pixel (spaxel 
\footnote{The term spaxel describes all pixels spanning the entire
  wavelength range of an IFS data cube that correspond to a single
  spatial pixel}) of the data cube, while rejecting outliers.  The
  result is a very high SNR estimate of the Airy pattern (and all
  wavelength dependent speckles) at each wavelength of the data
  cube. The faint companion represents a high frequency modulation and
  is thus excluded from the low order polynomial fit (as
  it represents a tilted line through the scaled data cube, it is
  present in any given spaxel over only a small range of
  wavelengths). The light from the parent star is then subtracted from
  each scaled slice of the data cube, and the result scaled back to
  the original grid, to form the result cube.

\begin{figure*}
\begin{center}
\includegraphics[width=\textwidth,angle=0]{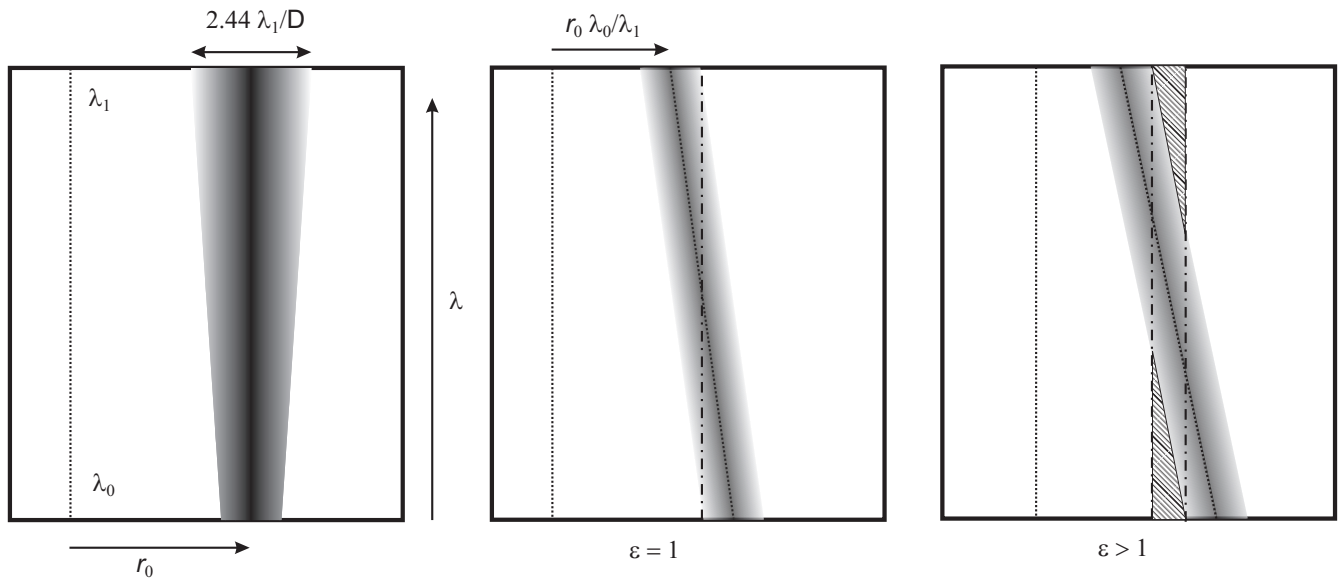}
\caption{The left hand panel is a schematic of a slice through the IFS
data cube, in the $x-\lambda$ plane, with the x axis along the line
joining the centre of the star with the centre of the faint
companion. The dotted line represents the peak of the stellar light
profile (as a function of lambda). The size of the companion,
quantified as $2.44 \lambda/D$, increases with wavelength. The middle
panel shows the data cube scaled inversely with wavelength, centred
on the star. The faint companion's separation is such that $\epsilon =
1$. The collapsed cube will be contaminated by companion light at all
$r$ values.  The right hand panel illustrates the situation for
$epsilon > 1$, where the hatched areas represent parts of the data
cube not contaminated by light from the companion. It is thus possible
to create a collapsed image free from companion light.}
\label{fig:epsilon}
\end{center}
\end{figure*}

The actual implementation of the concept for ground based
A.O. assisted integral field spectroscopic data (without a
coronagraph) has to account for low
Strehl that varies substantially with wavelength. Furthermore, we have
extended the original concept in two ways: (a) to extract spectra from
objects that lie close to the parent star and (b) optimise the signal
to noise ratio of the extracted spectra.  We elaborate on these
extensions in the following sub-sections. 

\subsection{Application to faint companions at small separations}

The SD technique utilises the wavelength dependence of PSF features to
obtain a high signal-to-noise estimate of the parent star PSF, while
rejecting the light from the faint companion.  We first derive the
limiting distance for which the technique is applicable in its
original proposed form, and then explain how we deal with close-in
companions.

\subsection{Bifurcation point for characterisation analysis}\label{bifurcation}

Consider a companion object located at radial distance $r$ (expressed
in angular units)  from the
primary object, imaged with a telescope of diameter $D$. The distance
to the first null of the Airy pattern is given by the usual
formulation
\begin{equation}
\Theta_0 = 1.22 {\lambda_0 \over D}
\end{equation}
where $\lambda_0$ is the shortest wavelength in the IFS data cube.
Defining the object extent as equal to 2\,$\Theta_0$, and the
wavelength range of the IFS as extending from $\lambda_0$ (shortest)
to $\lambda_1$ (longest), we obtain an expression for the movement of
the companion's centre in the scaled data cube as
\begin{equation}
\Delta r = r - r\,{\lambda_0 \over \lambda_1}  = r\,{\Delta\lambda  \over
  \lambda_1}
\end{equation}
Noting that the extent of the object stays constant in the scaled data
cube, we can then express the bifurcation point as 
\begin{equation}\label{eqn:bifurcation}
\Delta r = r\,{\Delta\lambda \over \lambda_1} = 2 \, \epsilon \times 1.22
\,{\lambda_0 \over D}
\end{equation}
where $\epsilon$ is a factor slightly greater than one.  The parameter
$\epsilon$ is explained in figure \ref{fig:epsilon}.

For $\epsilon
\equiv$ 1, the scaled data cube is entirely contaminated by light from
the companion, but only just so.  For $\epsilon >$ 1, a {\em clean}
collapsed image devoid of flux from the companion object can be made
by collapsing a small fraction of wavelength channels at either end of
the data cube. This is only necessary in the near vicinity of the
faint companion, as shown by the hatched areas in the right hand panel
of figure \ref{fig:epsilon}.  

The amount by which $\epsilon$ must exceed
unity depends on the SNR of the observation,
as sufficient number of wavelength channels must be used to form
the high SNR PSF estimate.  Table \ref{tab:bifurcation} lists
values of the bifurcation point for typical near-infrared band passes
for an 8 meter telescope, for a couple of typical values of
$\epsilon$.  It is obvious from the table that extended wavelength
coverage by the IFS is crucial for removing the flux from a close-in
companion in the scaled data cube. 

\begin{table}
\begin{center}
\caption{Bifurcation point in the data reduction as a function of
  bandpass of observations and factor $\epsilon$. $r$ is the value of
  the bifurcation radius in milli-arcseconds.}
\label{tab:bifurcation}
\begin{tabular}{lccccl}
\hline
Band & $\lambda_1$ & $\lambda_2$ & $\epsilon$ & $r$\\
 & $\mu$m & $\mu$m & & mas \\
\hline
\hline
H & 1.45 & 1.8 & 1.1 & 516\\
K & 1.95 & 2.45 & 1.1 & 661\\
H+K & 1.45 & 2.45 & 1.1 & 246\\
H & 1.45 & 1.8 & 1.2 & 563\\
K & 1.95 & 2.45 & 1.2 & 721\\
H+K & 1.45 & 2.45 & 1.2 & 268\\
\hline
\end{tabular}
\end{center}
\end{table}

\subsection{Extension to close-in faint companions}\label{iterative}

We have extended the SD technique to close-in faint
companions by using an iterative technique to remove the faint
companion's light from the PSF estimate made using the scaled data
cube.  Residual light from the faint companion in the estimated PSF
will reduce the flux level of the companion in the final result
cube. Furthermore, due to the wavelength scaling process, any residual
light will affect spectral channels differently, resulting in an error
in the continuum slope of the derived faint companion spectrum.
Section \ref{demo} shows that both the total flux and the continuum
slope were correctly measured for the observations reported here, thus
proving the efficacy of the proposed extension to the SD technique.

The following assumes that the position of the close-in faint
companion is already known (section \ref{results} mentions how this
might be derived from the IFS data cube itself). The companion is
also assumed to be an unresolved point source, with a full width at
half maximum (FWHM) roughly
equal to the diffraction limited FWHM for the telescope aperture in
question.  

Knowing both the position and the size of the source, we fit for the
companion amplitude and background level in individual spectral
channels of the observed data, prior to scaling. Subtracting an ideal
Airy pattern centred on the primary improves the robustness of the
fit, as does averaging $\sim$10 spectral channels.  Scatter in the fit
parameters is further reduced by describing each fit parameter with a
low order polynomial as a function of wavelength.  The goal is to
remove most of the flux from the faint companion, the quality of the
fit need not be very good.  The SD technique is then applied to the
data cube from which the faint companion light has been removed, to
first order.

The result data cube will contain residual companion flux due to
inaccuracies in the fit. However, as most of the PSF features will
have been eliminated by the spectral deconvolution process, we can use
the result data cube as a starting point for a second iteration,
fitting the faint companion more accurately than in the initial step. 
After subtracting the improved estimate of the faint companion light
from the observed data cube, we apply the SD
method again (second iteration) to obtain an improved result cube. Further
iterations are carried out until convergence is achieved, and the
extracted spectrum of the faint companion does not change. Fit
parameters such as companion position and width, fixed at first, can be
free parameters in subsequent iterations to improve the quality of the
fit. We applied this iterative technique to the data presented in the next
section, and found that only two iterations were required to achieve
convergence. 

The key to successfully applying the iterative procedure is to produce
a reasonable, if noisy, estimate of the faint companion flux in the
first iteration.  This can be achieved by subtracting an ideal Airy
pattern centred on the primary, or by subtracting data taken with the
same instrument, but with a different rotator angle. In either case,
the purpose is to estimate and remove the azimuthally symmetric
component of the PSF, so as to enhance the contrast at the location of
the faint companion.  This can also be achieved by fitting a radial
profile to the primary's light distribution, and subtracting it from
the observed data cube. The radial profile can also be used to model
the flux distribution from the faint companion, resulting in a better
subtraction of the companion flux in the data processing.

For faint companions located further than the bifurcation point,
applying the iterative procedure outlined above can still improve
the SNR of
the final extracted spectra, as it reduces the contamination from the
faint companion light in the PSF estimate.

\subsection{Optimising the SNR of the extracted spectra}
Observations carried out with a general purpose A.O. system suffer
from the added complication of moderate Strehl that varies
substantially with wavelength.  Consequently, the FWHM of the PSF does
not scale with $\lambda$ as expected for a purely diffraction limited
system. Instead, with Strehl increasing with wavelength, the PSF is
too broad at the shortest wavelengths, and too narrow at the longest
wavelengths, a feature which persists in the scaled data cube. This
results in a radial distance and wavelength dependent flux residual
after applying the SD technique. Consequently, the continuum of the
faint companion is incorrectly determined. To rectify these errors
caused by the varying Strehl, we fit and subtract a
radial profile for the primary light from each spectral channel of the
data cube prior to applying the SD technique. The data presented in
section \ref{demo} have been so treated. 

A further improvement in SNR of the extracted spectrum is obtained by
scaling the PSF estimate cube back to the original observed pixel
scale and subtracting it from the observed data cube, rather than
scaling the observed data cube and performing the subtraction of the
scaled cubes. Although the scaling is a linear operation, an
interpolation is performed to place all scaled channels on a common
grid, and this results in noise enhancement.  As the observed data
have much lower SNR (per spectral channel) than the PSF estimate
formed by the SD technique, the noise enhancement
is avoided almost completely by scaling and de-scaling the high SNR
PSF estimate rather than the lower SNR observed data cube.

\section{Demonstration}\label{demo}
We obtained observations of the local young
K-dwarf AB Doradus, which is known to have a M-dwarf companion (AB Dor
C) with $\Delta$K=5\,magnitudes only 0\farcs 2 away
\citep{close05}. AB Dor A has a
K-band magnitude of 4.6, making it an ideal AO guide star. Details of
the AB Dor C observations and data analysis are presented here as
proof-of-concept of the proposed extensions to the SD technique, as
described in previous sections.  It is also the first time the
SD method has been applied to a ground based
A.O. assisted IFS data set, with moderate Strehl and no coronagraph
present.  As such, it also serves as an experimental verification of
the SD method and
its applicability to ground based data sets. We also present a high SNR
H \& K band spectrum of AB Dor C extracted using this technique.
Further detailed analyses of the results of these observations are
published in the companion paper by \citet{Clo06} (Paper II) .

\subsection{Observations}

Data were obtained on the nights of 24th and 25th January 2006 with
the SINFONI instrument
\citep{sinfoni-thatte98,eisenhauer03,sinfoni-msgr04} at the Cassegrain
focus of the ESO VLT-UT4 (Yepun). Table~\ref{tab:obs} lists the
observations, all with the H$+$K grating. DIT is the time per
exposure, NDIT such exposures are averaged together by the readout
hardware, and the result is written to disk. The total exposure time
is made up of many such averaged frames.  In addition to the AB Dor
system, observations were made of a star in the Pleiades cluster to
serve as an M8 template of young age, and of several telluric
standards, at both 25 and 250 mas plate scales.

\begin{table}
\begin{center}
\caption{List of observations. All observations were made with the H+K
grating. The 250\,mas observations will not be discussed in this
paper.}
\label{tab:obs}
\begin{tabular}{lcccc}
\hline
Target & Spaxel Size & DIT & NDIT & Total exposure\\
 & (mas) & (s) & & (s) \\
\hline
\hline
AB Dor$^{\rm a}$ & 25 & 5.0 & 16 & 1280 \\
AB Dor$^{\rm b}$& 25 & 5.0 & 16 & 1280\\
AB Dor$^{\rm c}$& 250 & 0.83 & 4 & 6.64\\
Teide 1$^{\rm d}$& 250 & 300 & 1 & 600\\
\hline
\end{tabular}
\end{center}
{\bf $^a$}{-16.5$\deg$ PA}\\
{\bf $^b$}{+16.5$\deg$ PA}\\
{\bf $^c$}{These data were taken to bootstrap the spectral PSF
  from 250 mas pixel scale to the 25 mas pixel scale, if required}\\
{\bf $^d$}{Full designation is Cl* Melotte 22 Teide 1, template of M8
  spectral type}\\
\end{table}

AB Dor A was used as the guide star for the SINFONI adaptive optics
module \citep{bonnet04}.
The mean Strehl ratio
delivered was 0.37 on the 25th, and 0.35 on the 26th, as reported by
the SINFONI A.O. system itself. SINFONI was used in the 25\,mas scale
to ensure adequate sampling of the PSF. To enable removal of bad
pixels, and to sample SINFONI's rectangular spaxels, a dither pattern
with offsets of exactly $\pm$0.5 spaxels was used. The H+K grating was
used to maximise the wavelength coverage, resulting in a spectral
resolving power of $\sim$1500, and instantaneous wavelength coverage
of 1.4\micron -- 2.45\micron. No blank sky observations were taken
for AB Dor at the 25 mas pixel scale, as it was clear that the data
would be dominated by photon/speckle noise from AB Dor A.  The J band
was used for acquisition to avoid saturation of the detector array.
Seeing during the observations (as reported by the Atmospheric Site
Monitor) varied between 0\farcs 58 and 0\farcs 74, with a mean of
0\farcs 64 on the 25th, and between 0\farcs 61 and 0\farcs 86 with a
mean value of 0\farcs 69 on the 26th.

\begin{figure*}
\begin{center}
\includegraphics[width=0.4\textwidth,angle=90]{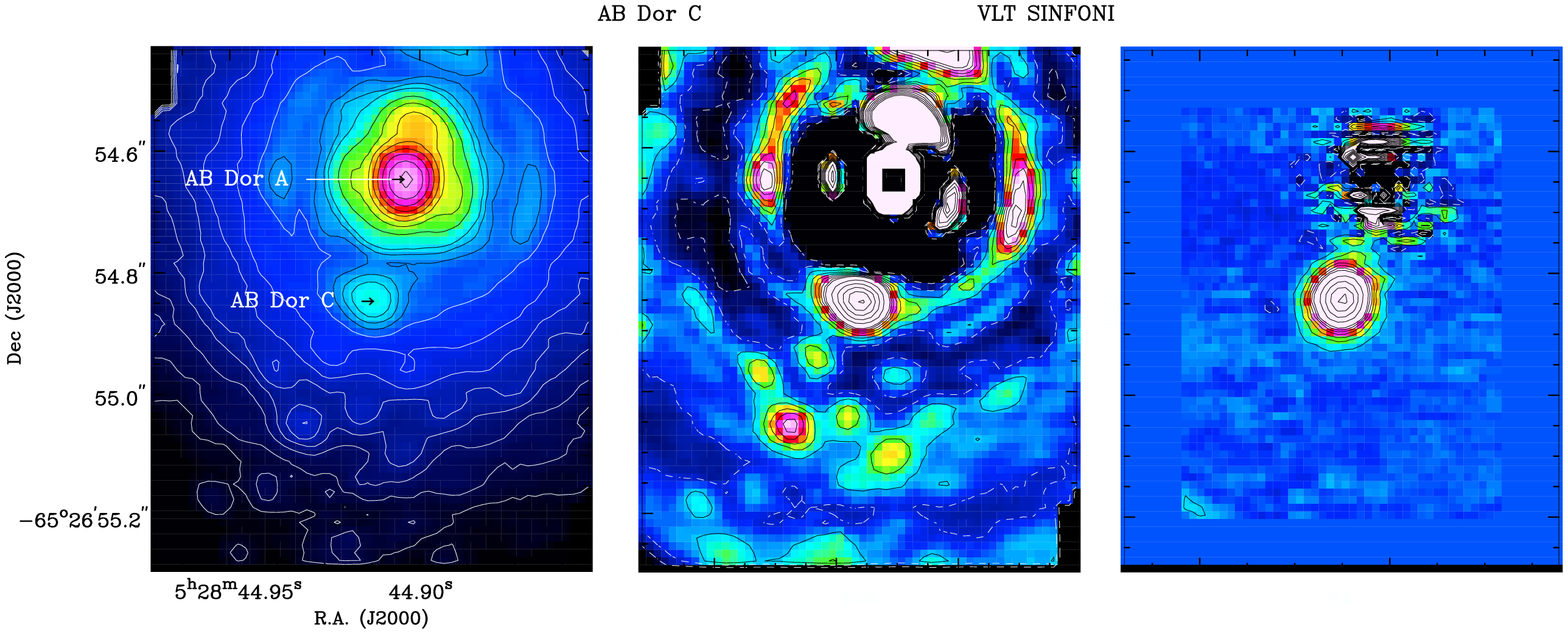}
\caption{Illustration showing the efficacy of the SD technique at
  removing both PSF artifacts and super-speckles from the SINFONI IFS
  data cube for the AB Dor system.  The left frame shows one
  wavelength slice of the observed data cube at 2.2 $\mu$m.  Note that
  the entire vertical extent of the image is only 0\farcs 9. The
  colour table is logarithmic (min 10$^1$, max 10$^4$). The contours
  are logarithmic, from 0.9 to 2.3 in steps of 0.1, and from 2.3 to
  4.0 in steps of 0.2.  The middle frame shows the same data, but with
  a radial profile fitted and subtracted, so as to highlight the PSF
  imperfections. The super-speckles are easily confused with real
  sources in this narrow-band image.  The colour table is now linear
  (min -10, max 25), with contours from -12.5 to 32.5 in steps of 5,
  and from 32.5 to 150 in steps of 20.  The four-fold symmetry of the
  Airy pattern arises from the superposition of the diffraction spikes
  from the secondary support structure on the Airy rings. The right
  hand frame shows the same wavelength slice of the data cube, after
  applying the SD technique iteratively. Colour table and contours are
  the same as for the middle frame. Super-speckles are completely
  absent at the lowest contour level of $\pm$2.5, corresponding to a
  1$\sigma$ error of $\leq$1 unit.}
\label{fig:channelmap}
\end{center}
\end{figure*}

\subsection{Data reduction}
Basic data reduction followed the standard SINFONI data reduction
procedure, outlined in \citet{schreiber04} \& \citet{abuter06}, with
extensions to deal with the 2K camera upgrade of February 2005
\citep{spiffi2k}.  The jitter pattern was corrected by combining data
cubes, one for each on-source exposure recorded, creating a single
mosaicked cube with a total observing time of 20 minutes for the data
of the 25th, and a second one (with different rotator angle) for the
26th. Accurate centroiding of AB Dor A in the resulting data cube
showed a residual variation with wavelength, implying that the
standard differential atmospheric refraction correction employed by
the pipeline reduction was not accurate enough.  Consequently, we
disabled the differential atmospheric refraction correction in the
standard reduction, and did a post-correction of the AB Dor A centroid
movement along the wavelength axis of the data cube, using a cubic polynomial
function.  The resulting improved model for differential atmospheric
refraction will be presented elsewhere.

It should be noted that even after disabling the differential
atmospheric refraction correction, there are still three
interpolations inherent in the production of a SINFONI data cube, two
of which use a nearest neighbour interpolation algorithm with a tanh
kernel, and one uses a polynomial interpolation algorithm. One
interpolation is required to ``straighten'' the spectra, a second does
re-gridding onto an uniform grid in wavelength, and a third corrects
for sub-pixel shifts between slices.  We have not evaluated the impact
of these interpolations in a quantitative manner, but it is likely
that they do impact the maximum contrast achieved by these
observations.

Implementation of the SD technique (and its extensions) used both IDL
and GIPSY data analysis packages, in addition to the routines from the
SINFONI pipeline.  GIPSY \citep{gipsy01} was used primarily for
visualisation of the data cubes with {\tt sliceview}, while most
computationally intensive tasks were performed in IDL.

\subsection{The SD technique and extensions}
AB Dor C is located 0\farcs 2 from AB Dor A, at P.A. 181$^\circ$ at the
epoch of these observations, based on orbit parameters determined by
\citet{nielsen05}.  Using equation \ref{eqn:bifurcation}, we note that
AB Dor C is closer than the bifurcation radius, thus requiring an
iterative procedure to identify the companion and extract its
spectrum.
 
The technique described in previous sections was applied to the AB Dor
data cubes to extract a spectrum of the companion AB Dor C.  As AB Dor
C is clearly visible in the raw data cube (see figure
\ref{fig:channelmap}), there was no problem in determining the
companion location. However, AB Dor C did overlap with a bright spot
in the diffraction pattern in the data taken on the 25th. The pattern
of four bright spots is caused by the telescope secondary mount
diffraction pattern superposed on the Airy pattern. Consequently,
accurate determination of the companion location did require a second
iteration of the SD technique, as explained in section \ref{iterative}
above.  Due to the strong dip in atmospheric transmission at the edges
of both H and K bands, the processing was confined to wavelength
ranges 1.457 to 1.80 $\mu$m and 1.95 to 2.45 $\mu$m respectively for
the H and K bands. A total of two iterations were needed for the SD
technique to yield a high SNR PSF estimate free from companion flux.

\subsection{Results}\label{results}

Figure \ref{fig:channelmap} shows one wavelength slice of the SINFONI
IFS data cube for the AB Dor system before and after the application
of the SD technique (and its extensions) described
above. The separation of AB Dor C from the parent was only 0\farcs 2
at this epoch, while the flux contrast was $\Delta$m = 4.90.  The left
hand frame shows the raw data, while the central frame has a smooth
radial profile fitted and subtracted from the raw data to highlight
the PSF artifacts, and in particular, the super-speckles. The right
hand frame shows the result of applying the technique described in
section \ref{concept}.  We can
successfully remove both Airy pattern residuals, as well as
super-speckles.  Note the four-fold symmetry of the Airy pattern, a
result of the superposition of the Airy ring with the diffraction from
the spider arms of the secondary. Unfortunately, AB Dor C is located
at the same orientation as one of the Airy peaks. We prefer to use
this data set as it has significantly better Strehl, although the
other data set (rotated by $33^\circ$) places AB Dor C at a more
favourable location relative to the Airy peaks.  Stepping through the
data cube along the wavelength axis, all PSF imperfections scale
linearly with wavelength, \footnote{mpeg movies of the raw, radial
profile subtracted and SD result data cubes can be viewed at {\tt
http://www-astro.physics.ox.ac.uk/$\sim$thatte/abdorc}} while the real
object stays at a fixed location. The data do demonstrate that the SD
technique can effectively separate companion and PSF contributions,
even if they are spatially coincident in a single wavelength channel.
The removal of the super-speckles is of particular value, as their
shape mimics point sources in the field in any narrowband image.

\subsubsection{Extracted spectra of AB Dor C}

\begin{figure}
\begin{center}
\includegraphics[width=0.5\textwidth,angle=0]{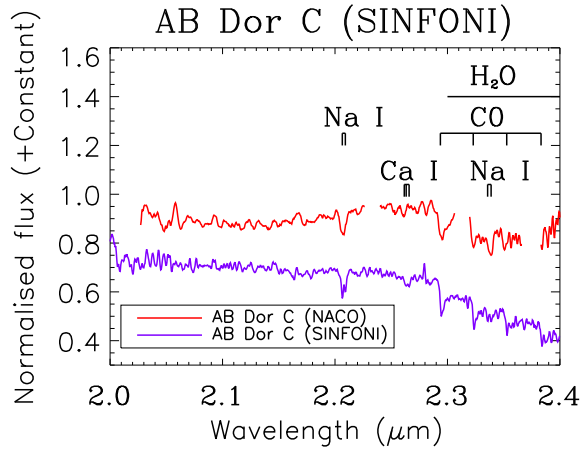}
\caption{ The K band extracted spectrum of AB Dor C, with some of the
prominent stellar features marked.  The K band spectrum obtained by
NACO A.O. long slit spectroscopy, at 0\farcs 150 separation
\citep{nielsen05} is also shown for comparison. Note that our spectrum
also correctly recovers the continuum slope, vital for deriving the
spectral type of this young object (see Paper II for details of the
spectral classification). The ``emission'' feature just shortward of
the first CO bandhead is a residual telluric feature that was not
correctly subtracted out.  }
\label{fig:KspecC}
\end{center}
\end{figure}

\begin{figure}
\begin{center}
\includegraphics[width=0.5\textwidth,angle=0]{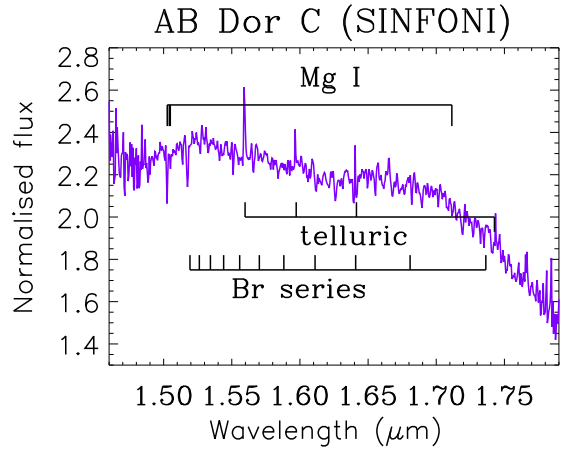}
\caption{
The H band extracted spectrum of AB Dor C, with some of the
stellar features marked.  Also marked are the location of the Brackett
series hydrogen recombination lines, as these were present in
absorption in the late B telluric standard.  Although every effort was
taken to accurately fit these lines, some residuals remain.  As no sky
observations were made, some of the brightest night sky OH emission
lines are also present in the derived spectrum.  The H and K spectra
were obtained simultaneously, by applying the SD technique (and its
extensions) to the data cube covering both H \& K bands. They are
shown separately for clarity.
}
\label{fig:HspecC}
\end{center}
\end{figure}

Figures \ref{fig:KspecC} and \ref{fig:HspecC} shows the normalised,
extracted spectra of AB Dor C, using the SD technique.  The data have
been smoothed with a boxcar of width 3 pixels, corresponding to 1.5
nm. The SNR of the K band spectrum exceeds 40, as derived from the
contrast values discussed in \ref{contrast}.  The quality of the
spectrum is limited by systematic errors in the division of telluric
features, as no sky observations were performed.  The NACO spectrum
previously obtained by \citet{close05} at a tighter 0\farcs 15
separation is also shown for comparison.  Note that the SD technique
is able to reproduce not only the spectral features but the continuum
slope as well, in contrast to A.O. long-slit measurements with varying
slit filling factors. The output data cube, after applying the SD
technique, is used to measure the position, flux, and spectral type of
AB Dor C. At the epoch of these observations, AB Dor C was located
201.7$\pm$10 milli-arcseconds from AB Dor A, at
P.A. 180.78$^\circ$. The K magnitude was 4.90 mag fainter than AB Dor
A, which, combined with the 2MASS measurement for AB Dor A yields
K=9.59 for AB Dor C.  A spectral type of M5.5$\pm$0.5 can be derived
by comparing the observed spectrum with young and old templates.  A
detailed analysis of the properties of AB Dor C (including its
spectral type determination) is presented in Paper II.
Note that both the companion flux and its spectral continuum slope are
correctly derived using the iterative method described in section
\ref{iterative} above. 

\subsubsection{Achieved Contrast}\label{contrast}

\begin{figure}
\begin{center}
\includegraphics[width=0.5\textwidth,angle=0]{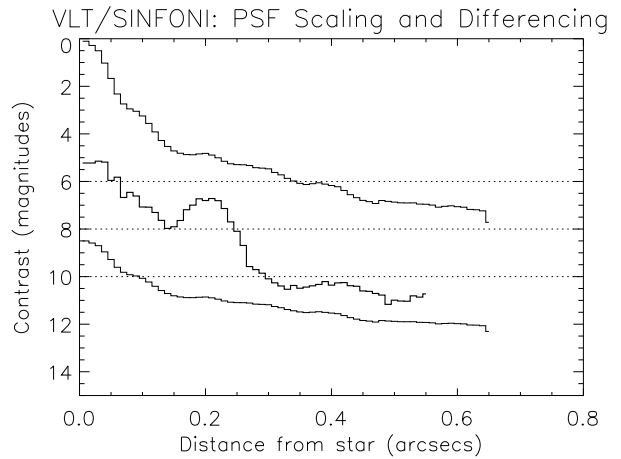}
\caption{ The plot shows three curves - the top curve is the radial
profile of AB Dor A.  The bottom curve is the square root of the top
curve, so it represents the photon noise limit. The middle curve is
the standard deviation in the SD result frame, so it is a measure of the
residual noise. The horizontal lines at 6, 8 and 10 magnitudes are for
reference. The large bump at 0\farcs 2 in the SD result frame is due
to AB Dor C. The achieved contrast is 9 mag at 0\farcs 2, and 11 mag
at 0\farcs 5, in 20 mins exp. time at R$_{\rm eff}$=400. As no
coronagraph is used, very high contrasts can be obtained at small
inner working radii.
}
\label{fig:contrast}
\end{center}
\end{figure}
  
The remarkable success of the SD technique at
removing light contamination resulting from PSF imperfections is
illustrated in figure \ref{fig:contrast}. The plot shows the resulting
radial profile for a 20 min exposure of the AB Dor system observed
with SINFONI, with a spectral width of 5.5 nm, using the SD 
technique. AB Dor C is clearly visible as a bump in the radial profile
at 0\farcs 2.  The curve shows the contrast achieved (1$\sigma$) is
$\sim$ 9 mag at 0\farcs 2, and 11 mag at 0\farcs 5, without using a
coronagraph.  Note that the effective spectral resolution (R=400) is
higher than that achieved by SDI (R=50) \citep{lenzen04,biller06}.
The achieved contrast is still $\sim$1 mag from the photon noise
limit, although we suspect that some residual noise is attributable to
non-optimal interpolations in the data reduction procedure, and can be
improved upon.  The ultimate limitations of the technique will be
addressed in a separate paper.

\subsection{Faint object detection}

AB Dor C is only 5 mag fainter than AB Dor A, and could thus be easily
detected in the raw data cube. However, we note that the SD technique
can also be used for detection of faint companions.  Any feature whose
radial distance from the primary does not scale with wavelength
generates a radial streak in a collapsed image created by averaging
the scaled data cube along the spectral dimension. Subtracting the
collapsed image (scaled back appropriately for each wavelength) from
the original data cube results in a diminished companion flux plus an
inward negative streak at the shortest wavelengths, a diminished flux
companion plus a symmetric negative radial streak at central
wavelengths, and a weaker companion plus an outward negative streak at
the longest wavelengths.  The final collapsed image shows the
characteristic pattern of a positive object sitting atop a long
negative radial streak centred on the object (\citep[see both
``planets'' in fig. 26 of][]{sparks02}). This characteristic pattern
can be used to detect faint companions.  Furthermore, if the primary's
radial light profile is fitted and subtracted from each wavelength
slice of the data cube prior to applying the SD technique, the
contrast of the faint companion can be further enhanced.

If the technique is applied to data cubes taken at two different rotator
angles and subtracted from each other prior to step 1, only features
that are stationary on the sky will remain. In addition to showing the
characteristic pattern described above, they will also show a positive
and negative image of true celestial objects, separated by the
difference in rotator angles.  The resulting pattern can be easily
used to identify potential faint companions.

\section{Conclusions}\label{conclusion}

In this paper we have introduced extensions to the SD technique
proposed by \citet{sparks02} for achieving high contrast imaging
spectroscopy with an AO-fed integral field spectrograph at small inner
working radii.  Applying this technique to real data, we have shown it
to provide very high contrast spectra of faint companions very
close to bright stars.  The achieved contrast is substantially better
than previously achieved by other techniques, especially as no
coronagraph was used.  The absence of a coronagraph removes limits on
the smallest inner working radius, especially valuable as most
exo-planet candidates are expected to lie at very small separations
from the parent star.  The SD technique holds great promise for
direct imaging of exo-planets, as it simultaneously detects and
spectrally characterises any faint companion, thus removing the need
for expensive and time-consuming follow-up observations, either to
detect common proper motion or to obtain a spectrum of the faint
companion.  In addition, the technique does not require any
assumptions about the companion's spectral characteristics
(e.g. presence of a CH$_4$ feature), and can therefore be applied to
any high contrast application.

Our demonstration of the efficacy of the SD technique to obtain
spectra of the close-in faint companion AB Dor C shows that image
slicer based integral field spectrographs are capable of achieving
very high contrasts, contrary to the expectations of several
groups \citep{cheops-ifs,epics06,berton06} who were concerned that
large non-common-path errors would limit the contrast achievable with
these systems.  The SINFONI IFS design \citep{tecza00,tecza03}
exclusively uses classically polished flat glass mirrors in its image
slicer, thus achieving very small non common path errors, as
demonstrated here. Indeed, we have shown that an integral field
spectrograph alone can provide a large fraction of the total contrast
requirement of an exo-planet direct detection instrument.

Successful application of the SD technique requires large instantaneous
wavelength coverage, strongly favouring image slicer based IFS
designs.  These naturally provide large simultaneous wavelength
coverage, while it is rather difficult to achieve a large bandwidth in
lenslet array based designs.  The requirement to keep non common path
errors to an absolute minimum strongly favours image slicer designs
with classically polished glass slicing optics (not just slicing
mirrors, but the entire slicer optics), as implemented in the SINFONI
spectrograph \citep{tecza00,tecza03}.

\section{Acknowledgements}

We thank the ESO Director-General for allocating Director's
discretionary time to carry out these observations.  We thank MPE for
use of the {\tt spred} SINFONI data reduction software. This paper is 
based on observations collected at the European Southern Observatory,
Chile under ESO programme ID 276.C-5013. NT, FJC \& MT are funded
through Marie-Curie Excellence Grant MEXT-CT-2003-002792. EN is
supported by a Michelson Fellowship. LMC is supported by an NSF CAREER
award and the NASA Origins of Solar Systems program. We thank the
referee, W.B.Sparks for extensive comments that greatly improved the
paper.

\bibliographystyle{mn2e}
\bibliography{thatte_v3}

\label{lastpage}
\end{document}